\documentclass[pra,aps,twocolumn,superscriptaddress,showpacs,amsmath,amssymb,floatfix]{revtex4}

\usepackage{graphicx}

\newcommand{\be}{\begin{equation}}
\newcommand{\ee}{\end{equation}}

\begin{document}
\title{Two-body problem in periodic potentials}
\author{M. Wouters}
\affiliation{BEC-INFM and Dipartimento di Fisica, Universita' di Trento, 
1-38050 Povo, Italy}
\affiliation{TFVS, Universiteit Antwerpen, Universiteitsplein 1,
2610 Antwerpen, Belgium}
\author{G. Orso}
\affiliation{BEC-INFM and Dipartimento di Fisica, Universita' di Trento, 
1-38050 Povo, Italy}

\begin{abstract}

We investigate the problem of two atoms interacting via a short range $s$-wave potential 
in the presence of a deep optical lattice of arbitrary dimension $D$.
Using a tight binding approach, we derive analytical results for the
properties of the bound state and the scattering amplitude.
We show that the tunneling through the barriers induces a dimensional crossover
from a confined regime at high energy to an anisotropic three dimensional regime at low
energy. 
The critical value of the scattering length needed to form a two-body bound state
shows a logaritmic dependence on the tunneling rate for $D=1$ and a power law
for $D>1$.
For the special case $D=1$, we also compare our analytical predictions with exact numerics,
finding remarkably good agreement.

\end{abstract}

\maketitle

\section{Introduction}
Recent progress in controlling and manipulating the interatomic  
forces via Feshbach resonances \cite{ketterle} and the availability
of tunable periodic potentials generated by laser beams \cite{varenna} 
are opening new fronts in the research on ultracold atomic gases.
An important field of activities concerns the study of two-body physics
in the presence of an external potential. 
This two-body problem is directly related to current experiments and moreover, 
it forms a basic step towards understanding  many body states where
strong correlations can play a significant role.

From the experimental point of view, a major achievement has been 
the formation of
weakly bound diatomic molecules in experiments  with two-component clouds of 
fermionic atoms \cite{jila,mit,Innsbruck,ens}. 
Recently, also the binding energy of confinement induced 
molecules  has been measured by 
the group of Esslinger \cite{esslinger} working with samples of 
potassium $^{40}$ K in a deep two dimensional optical lattice.

From the theoretical front, the two-body problem
in the presence of harmonic trapping,
has been solved analytically both for
$s$-wave \cite{gora,olshanii,busch,sbiscek} and $p$-wave short range 
interactions \cite{granger,sbiscekp}. 
The case of a periodic external potential is even more
interesting as the center of mass and the relative 
motion no longer decouple \cite{orso,mora}. 
This gives rise to an interesting interplay between 
tunneling and confinement effects. 

The properties of the bound state  of two interacting atoms
in the presence of a one dimensional  optical lattice have been 
investigated {\sl numerically} in Ref.\cite{orso}. 
For sufficiently deep lattices and binding energy $E_b$ small  compared to the
interband gap, two different regimes were identified:
a  quasi {\sl two}-dimensional regime for $|E_b| \gg t$, $t$ being the interwell tunneling rate,
where the molecular wavefunction is localized in one well and
an  anisotropic {\sl three} dimensional regime for $|E_b|\ll t$
where the wavefunction spreads over many lattice sites \cite{fedichev}.  
An analogous dimensional crossover occurs for the scattering
amplitude as the total energy of the two incident states becomes 
large compared to the tunneling rate \cite{tc}. 

In this work we apply the tight binding approach introduced in Ref.\cite{tc} to study 
analytically the properties of the bound state and the 
scattering amplitude of two interacting atoms in the presence 
of a $D$ dimensional tight optical lattice. We derive simple yet 
accurate predictions for many quantities that can be directly 
measured in current experiments with dilute ultracold gases. 

The paper is organised as follows. In Section II we present the general
theory to solve the two-body problem in the presence of a periodic potential.
In section III we apply the tight binding approach to the case $D=1$
and we extensively compare our analytical results with the exact numerics of Ref.\cite{orso}. 
In section IV we extend our analysis to the case of optical lattices of higher ($D=2,3$)
dimension.

\section{GENERAL THEORY}
We consider two atoms in ordinary space interacting in the presence of a 
$D$ dimensional periodic potential 
\be\label{vopt1}
V_{opt}(\mathbf{x})=s E_R \sum_{i=1}^D \sin^2\left(\frac{\pi x_i}{d}\right)
\ee 
where $D=1,2,3$. Here $s$ is the 
intensity of the laser beams generating the optical lattice, 
$E_R=\hbar^2 \pi^2/2m d^2 $ is the recoil energy, 
$d$ is the lattice period and $m$ the 
atom mass. There is no confinement in the remaining $3-D$ directions.

Modelling the interatomic interaction by a $s$-wave pseudopotential 
with coupling 
constant $g=4\pi \hbar^2 a/m$, with $a$ the 3D scattering length, 
the Schrodinger equation takes the form
\begin{eqnarray}
  \left(-\frac{\hbar^2}{m}\nabla^{2}_\mathbf{r}-\frac{\hbar^2}{4m}\nabla^2_\mathbf{R}+
  V(\mathbf{R},\mathbf{r_\parallel}) 
  +g\delta(\mathbf{r})\frac{\partial }{\partial r} r\right)
  \Psi=E\Psi,
\label{schrodinger}
\end{eqnarray}
where $\mathbf{r}=\mathbf x_1-\mathbf x_2$ is the relative distance between
the two atoms, $\mathbf R$ and $\mathbf r_\parallel$ are the components of the 
center of 
mass and the relative coordinates in the confined directions and 
$V(\mathbf{R},\mathbf{r_\parallel})=V_{opt}(\mathbf{x}_1)+
V_{opt}(\mathbf{x}_2)$ is the total external field. 

With respect to the case of harmonic trapping, 
the periodic field 
introduces a conceptually new difficulty  
related to the fact that the center of mass and the 
relative motion no longer separate.  
Nevertheless, the quasi-momentum $\mathbf Q$, associated to the center of mass 
variable $\mathbf R$, remains a conserved quantity even in the 
presence of interaction and can be used to 
classify the solutions of Eq.(\ref{schrodinger}). This can be seen by 
noticing that Eq.(\ref{schrodinger}) is invariant under the transformation
\begin{eqnarray}
&\mathbf R &\rightarrow \mathbf R+\mathbf G\\
&\mathbf r &\rightarrow \mathbf r
\end{eqnarray}
which shifts the center of mass of the two atoms by a lattice vector 
$\mathbf G$ leaving unchanged their relative distance.

\subsection{Bound state}
Bound states are solutions of Eq.(\ref{schrodinger}) whose energy $E$ does not belong to the energy
spectrum in the absence of interaction. In this case Eq.(\ref{schrodinger}) can be written in the
integral form
\begin{equation}
  \Psi(\mathbf{r},\mathbf{R})=\int d\mathbf{R}^{\prime} G_E(\mathbf{r},\mathbf{R};\mathbf{0},\mathbf{R}^{\prime}) 
  g \frac{\partial }{\partial r^\prime} \left(r^\prime 
\Psi(\mathbf{r}^{\prime},
  \mathbf{R}^{\prime})\right)_{\mathbf{r}^\prime=0},
  \label{schrodinger2}
\end{equation}
where $G_E$ is the Green function associated to Eq.(\ref{schrodinger}) with $g=0$. 
The behaviour of the Green function at short distance is dominated by the kinetic term in 
Eq.(\ref{schrodinger}).
Taking into account that for small $r$
\begin{eqnarray}
\int \frac{e^{i \mathbf Q (\mathbf R-\mathbf R^\prime)} 
e^{i \mathbf k \mathbf r}}
{\hbar^2 k^2/m +\hbar^2 Q^2/4m} \frac{d^3\mathbf k}{(2\pi)^3}
\frac{d^D\mathbf Q}{(2\pi)^D} \notag \\
= \frac{m}{4\pi\hbar^2 r}\int 
e^{i \mathbf Q (\mathbf R-\mathbf R^\prime)}\frac{d^D\mathbf Q}{(2\pi)^D},
\end{eqnarray}
the Green function admits the following expansion
\begin{equation}
  G_E(\mathbf{r},\mathbf{R};\mathbf{0},\mathbf{R}^{\prime})
 = -\frac{m}{4\pi \hbar^2 r}\delta^D\left( \mathbf{R}-\mathbf{R}^{\prime }\right)
+K_{E}\left(\mathbf{R},\mathbf{R}^{\prime }\right)+O(r),  \label{defBC}
\end{equation}
where $K_{E}\left(\mathbf{R},\mathbf{R}^{\prime }\right)$ is a regular kernel which depends 
on the energy and the external potential. 

When inserted into Eq.(\ref{schrodinger2}), Eq.(\ref{defBC}) yields the Bethe-Peierls 
boundary condition 
$\Psi(\mathbf{r},\mathbf{R})\sim (a/r-1)f(\mathbf{R})$ for 
$r\rightarrow 0$,  
where $f(\mathbf{R})$ is a function of
the center of mass position satisfying the integral equation
\begin{eqnarray}
\frac{1}{g}f\left( \mathbf{R}\right)  &=&\int d \mathbf{R}^{\prime }~
K_{E}\left(
 \mathbf{R}, \mathbf{R}^{\prime }\right) f\left(  \mathbf{R}^{\prime }\right).   \label{int2}
\end{eqnarray}
The solution of Eq.(\ref{int2}) yields the energy $E$ as a 
function of the 3D scattering length $a$. 

Two comments are in order here.
{\sl First}, the integral equation 
(\ref{int2}) is completely general and is valid for any external potential. In the special case of 
harmonic trapping, $f(\mathbf{R})$ is also an eigenstate of the center of mass Hamiltonian and 
Eq.(\ref{defBC}) reduces to a simpler {\sl algebraic} equation.
{\sl Second}, the conservation of quasi-momentum in periodic potentials is ensured by the
symmetry property  
$K_E( \mathbf{R}, \mathbf{R}^\prime)=
K_E( \mathbf{R}+\mathbf{G}, \mathbf{R}^{\prime }+ \mathbf{G})$ of the kernel, 
holding for any lattice vector $ \mathbf{G}$.

In the following we are mainly interested in the weakly bound state whose energy $E$ is below the
minimum $E_{ref}$ of the non interacting energy spectrum \cite{note3D}. 
For a given quasi-momentum $\mathbf Q$,
the lowest energy solution of Eq.(\ref{schrodinger}) for $g=0$ corresponds to  
the case where both atoms occupy the state of quasi-momentum $\mathbf Q/2$ in the lowest
Bloch band, that is $E_{ref}(\mathbf Q)=2\epsilon_1(\mathbf Q/2)$.
We define the binding energy of the molecule as $E_b\equiv E-E_{ref}<0$.

A direct consequence of the non separability of the center of mass and 
the relative motion in the lattice is that the binding energy 
becomes a function of the quasi-momentum of the molecule, or, the other way around, 
the tunneling properties of the molecule (strongly) depend on the value of the scattering length.
A second important difference from the case of harmonic trapping is that, in the lattice,
Eq.(\ref{int2}) predicts a {\sl finite} critical value of the scattering length
$a=a_{cr}<0$ such that no bound state exists for $a_{cr}<a<0$. 
Notice that in free space $(s=0)$, a bound state exists only for positive
value of the scattering length, i.e. $a_{cr}=-\infty$.

\subsection{Scattering amplitude}
Let us consider a non interacting two-particel state 
$\Psi_0(\mathbf{r},\mathbf{R})=\phi_{\mathbf{n_1},\mathbf{q}_1}(\mathbf{r}_1)
\phi_{\mathbf{n_2},\mathbf{q}_2}(\mathbf{r}_2)e^{i \mathbf k_\perp \mathbf 
r_\perp}$, where  $\phi_{\mathbf{n},\mathbf{q}}(\mathbf{x})=
\prod_{j=1}^D u_{n_j,q_j}(\mathbf{x})$. Here
$u_{n,q}(z)$ are the solutions of the one dimensional Hamiltonian 
$H=-(\hbar^2/2m)d^2/dz^2+V_{opt}(z)$ with energy $\varepsilon_{n}(q)$, 
with $n$ the band index and $q$ the quasi-momentum.
A quantity of great physical interest is the scattering amplitude associated 
to $\Psi_0$, the incident state. 
In the presence of the lattice, we {\sl define} it as 
\be\label{scatt}
f_{sc}[\Psi_0]=a \int \Psi_0^*(0,\mathbf{R})  \partial_r \left (
r \Psi(\mathbf r,\mathbf R)\right )_{r=0} d\mathbf R,
\ee
where $a$ is the 3D scattering length and $\Psi(\mathbf r,\mathbf R)$
is the solution of Eq.(\ref{schrodinger}), which can be written as
\begin{eqnarray}
&& \Psi(\mathbf{r},\mathbf{R})=\Psi_0(\mathbf{r},\mathbf{R}) \label{schrodinger3}\\
&&+\int d\mathbf{R}^{\prime} G_{E+i0}(\mathbf{r},\mathbf{R};\mathbf{0},\mathbf{R}^{\prime}) 
  g \frac{\partial }{\partial r^\prime} \left(r^\prime \Psi(\mathbf{r}^{\prime},
  \mathbf{R}^{\prime})\right)_{\mathbf{r}^\prime=0}.\notag
\end{eqnarray} 
Here $G_{E+i0}$ is the {\sl retarded} Green function  
evaluated at the energy $E=\epsilon_{\mathbf n_1}(\mathbf q_1)+\epsilon_{\mathbf n_2}(\mathbf q_2)
+\hbar^2 k_\perp^2/m$ of the incident state and $\epsilon_{\mathbf n}(\mathbf q)=
\sum_{j=1}^D \epsilon_{n_j}(q_j)$.
 Introducing
$f(\mathbf R)=\partial_r \left (
r \Psi(\mathbf r,\mathbf R)\right )_{r=0}$ and making use of the
expansion (\ref{defBC}) in Eq.(\ref{schrodinger3}), we find
the {\sl inhomogeneous} integral equation
\begin{eqnarray}
f\left( \mathbf{R}\right)  &=& \Psi_0(\mathbf{R},\mathbf{0}) 
+g \int d\mathbf{R}^{\prime }~K_{E+i0}\left(\mathbf{R},\mathbf{R}^{\prime }
\right) 
f\left( \mathbf{R}^{\prime }\right).\notag\\
\label{nonomo}  
\end{eqnarray}
which has to be solved to find the scattering amplitude (\ref{scatt}).
The comments below Eq.(\ref{int2}) applies here as well.

\subsection{Numerical and tight binding solution}
Equations (\ref{int2}) and (\ref{nonomo}) can be solved numerically 
following the method developed in Ref.\cite{orso} for the case of a 1D optical lattice. 
In order to extract the regular kernel $K_E$ from the Green function, 
the latter is expanded in the basis of the non interacting states  
\begin{eqnarray}
 && G_E(\mathbf{r},\mathbf{R};\mathbf{0},\mathbf{R}^{\prime}) = \label{GEexplicit}\\
 &&\sum_{\mathbf{n}_1,\mathbf{n}_2} \int\frac{d^{3-D}\mathbf{k}_{\perp }}{\left( 2\pi \right) ^{3-D}}
\int_{-q_B}^{q_B}\frac{d^D \mathbf{q}_1}{(2\pi)^D}\frac{d^D \mathbf{q}_2}{(2\pi)^D} 
 e^{i\mathbf{k_\perp}\cdot \mathbf{r_\perp}}\notag\\ 
   && \frac{   \phi_{\mathbf{n}_1,\mathbf{q}_1}(\mathbf{R}+\frac{\mathbf{r_\parallel}}{2})
    \phi_{\mathbf{n}_2,\mathbf{q}_2}(\mathbf{R}-\frac{\mathbf{r_\parallel}}{2})
\phi_{\mathbf{n}_1,\mathbf{q}_1}^*(\mathbf{R}^\prime)\phi_{\mathbf{n}_2,\mathbf{q}_2}^*
(\mathbf{R}^\prime)}
  {E-\epsilon_{\mathbf{n}_1}(\mathbf{q}_1)-\epsilon_{\mathbf{n}_2}(\mathbf{q}_2)-\hbar^2k_\perp^2/m}.
\notag
\end{eqnarray}
Since the singular term in the rhs of Eq.(\ref{defBC}) does not depend
on the external potential, we add and substract from $G_E$,
the Green function (\ref{GEexplicit}) evaluated for $s=0$. This 
permits us to write the kernel as
$K_E(\mathbf R,\mathbf R^{\prime })=\lim_{r\rightarrow 0}
[G_E(\mathbf{r},\mathbf R;\mathbf{0},\mathbf R^{\prime})-
G_E^{s=0}(\mathbf{r},\mathbf R;\mathbf{0},\mathbf R^{\prime})]+
K_E^{s=0}(\mathbf R,\mathbf R^{\prime })$, where the latter is given by
\be\label{GEfree}
K_E^{s=0}(\mathbf R,\mathbf R^{\prime })=\frac{m}{4\pi \hbar^2}
\int \frac{d\mathbf P}{(2\pi)^D} e^{i \mathbf P(\mathbf R-\mathbf R^{\prime })}
\sqrt{|E|+\hbar^2P^2/4m}.
\ee
The term in square brackets is evaluated numerically from Eq.(\ref{GEexplicit}).
Once the kernel $K_E$ is known,  Eqs (\ref{int2}) and 
(\ref{nonomo}) can be solved using standard diagonalization routines. This numerical method has been
applied in Ref.\cite{orso} for the case $D=1$. 
It should be noticed that for $D>1$ the numerical effort 
needed to solve Eqs (\ref{int2}) and (\ref{nonomo}), using the above method,
grows considerably. 

In this paper we restrict ourselves to the case of sufficiently 
deep lattices and energy $E$ small compared to the interband gap $\epsilon_g$.
In this limit the  energy dependence of the 
Green function in Eq.(\ref{GEexplicit}) comes {\sl entirely} from 
the lowest Bloch band and the two-body problem can be solved perturbatively 
using a tight binding approach pioneered in Ref.\cite{tc}.

\section{ONE DIMENSIONAL LATTICE}
Below, we discuss in details the case of a one dimensional 
optical lattice $(D=1)$ by comparing our analytical predictions with the
exact numerics of Ref.\cite{orso}. 
\subsection{Bound State} 
We start our analysis by noticing that for $D=1$ the integration over $\mathbf k_\perp$ 
in Eq.(\ref{GEexplicit}) is ultraviolet divergent  for $\mathbf r_\perp \rightarrow 0$. 
Taking into account that the singular term in the rhs of 
Eq.(\ref{defBC}) does not depend on energy,  
we can write identically:
\begin{eqnarray}
K_E(Z,Z^{\prime })&=&[G_E(\mathbf{r},Z;\mathbf{0},Z^{\prime})
-G_{E=E_{ref}}(\mathbf{r},Z;\mathbf{0},Z^{\prime})]_{r=0}\notag \\
&+&K_{E=E_{ref}}(Z,Z^\prime).
\end{eqnarray}
We consider first the term between square brackets
which contains the energy dependent part of the kernel. 
The integration over $\mathbf k_\perp$ in Eq.(\ref{GEexplicit}) 
converges even for $\mathbf r_\perp =0$. 
Moreover, for energy $E$ small compared to the
interband gap, only the lowest band $n_1=n_2=1$ contribute
significantly.
For sufficiently large values of the laser intensity, 
the states of the lowest Bloch band
can be written in terms of Wannier functions as 
$\phi_{1q_z}(z)\sim \sum_{\ell}e^{i\ell q_z d}w(z-\ell d)$, where 
\be\label{gauss}
w(z)=\frac{1}{\pi^{1/4}\sigma^{1/2}}\exp\left(-\frac{z^2}{2\sigma^2}\right)
\ee
is a variational gaussian ansatz. By minimizing the energy 
$\int dz w(z)[-(\hbar^2/2m) d^2/dz^2+V_{opt}(z)] w(z)$ with respect to 
$\sigma$, one finds  $d/\sigma\simeq \pi s^{1/4}\exp(-1/4\sqrt s)$.

Substituting the tight binding expression for the Bloch states
in Eq.(\ref{GEexplicit}), we 
immediately see that the term in square brackets is diagonalized
by the ansatz 
\be\label{ansatz}
f_Q(Z)\sim \sum_{\ell} w^2(Z-\ell d)e^{iQZ},
\ee 
where $Q$  is the quasi-momentum of the molecule.
Motivated by this fact, we have verified numerically that 
the ansatz (\ref{ansatz}) is a solution of Eq.(\ref{int2})
with $E=E_{ref}$, that is
\be
\int dZ^\prime K_{E=E_{ref}}(Z,Z^\prime)f_Q(Z^\prime)dZ^\prime=
f_Q(Z) m/4\pi\hbar^2 a_{cr},
\ee 
where $a_{cr}=a_{cr}(Q)$ is the critical value of the scattering length 
needed to form a two-body bound state with quasi-momentum $Q$. The ratio $d/a_{cr}$ 
as a function of the laser intensity has been evaluated numerically
in Ref.\cite{orso} and in Ref.\cite{tc} an analytical estimate was derived
for the special case $Q=0$.

By inserting the ansatz (\ref{ansatz}) into Eq.(\ref{int2}) and integrating 
over $Z^\prime$, we obtain
\be\label{eb-1d}
\left(\frac{1}{a}-\frac{1}{a_{cr}}\right)\sqrt{2 \pi}\sigma
=-\alpha_{1D}(E_b,Q),
\ee
where 
\begin{eqnarray}
&&\alpha_{1D}(E_b,Q)=\label{a-1D} \\
&&-\int_{-\pi}^{\pi}\frac{dq}{2\pi} 
\ln \left[\frac{-E_b-2\epsilon_1(Q/2)+\epsilon_1(q)+\epsilon_1(Q-q)}
{\epsilon_1(q)+\epsilon_1(Q-q)}
\right]\notag
\end{eqnarray}
and we have used the definition of binding energy $E_b \equiv
E-2 \epsilon_1(Q/2)$.
Equations (\ref{eb-1d}-\ref{a-1D}) give $E_b$ as a 
function of the quasi-momentum $Q$, the scattering length $a$
and the lattice parameters.  

Let us first discuss the case of zero quasi-momentum. 
By substituting the tight binding dispersion $\epsilon_1(q)=2t(1-\cos(qd))$
in Eq.(\ref{a-1D}) and integrating over $q$, we obtain 
\be\label{aba}
\alpha_{1D}(E_b,0)=\ln\left (1+\frac{|E_b|}{4t}+\sqrt{\frac{|E_b|}{2t}
+\left(\frac{E_b}{4t}\right)^2}\right ).
\ee
Equation (\ref{aba}) describes the dimensional crossover in the binding
energy as the ratio $|E_b|/4t$ is varied.
The  three dimensional regime $|E_b|\ll 4t$ is charachterized by a 
molecular wavefunction extended over many lattice sites.
Expanding Eq.(\ref{aba}) around $E_b=0$ and inserted it into 
Eq.(\ref{eb-1d}), we find  
\be\label{gen} 
1/a-1/a_{cr}=C \sqrt{|E_b| m^*}/\hbar,
\ee
 where
$m^*=\hbar^2/2 t d^2$ is the atomic effective mass evaluated 
at the bottom of the band and $C=d/\sqrt{2\pi} \sigma$.
The above result shows that the optical lattice gives rise to
an effective shift of the resonance from $1/a=0$ to
$1/a=1/a_{cr}<0$. 
In the quasi 2D regime $4t \ll |E_b|\ll \epsilon_g$, the two interacting atoms
are localized at the bottom of the same optical well, where, 
to a first approximation,  the potential (\ref{vopt1}) 
is harmonic with frequency $\omega_{0}=\hbar/m\sigma^2$.
In this case, the center of mass and the relative motion 
decouple and the problem can been solved analytically. 
In the relevant limit $|E_b|\ll \omega_0 \sim \epsilon_g$, the binding energy is given by
\be\label{ho}
E_b^{ho}=-\lambda \hbar \omega_0 \exp(-\sqrt{2\pi}\sigma/|a|)
\ee
with $\lambda=0.915/\pi$ \cite{gora}.  
\begin{figure}[tb]
\begin{center}
\includegraphics[width=\columnwidth,angle=0,clip]{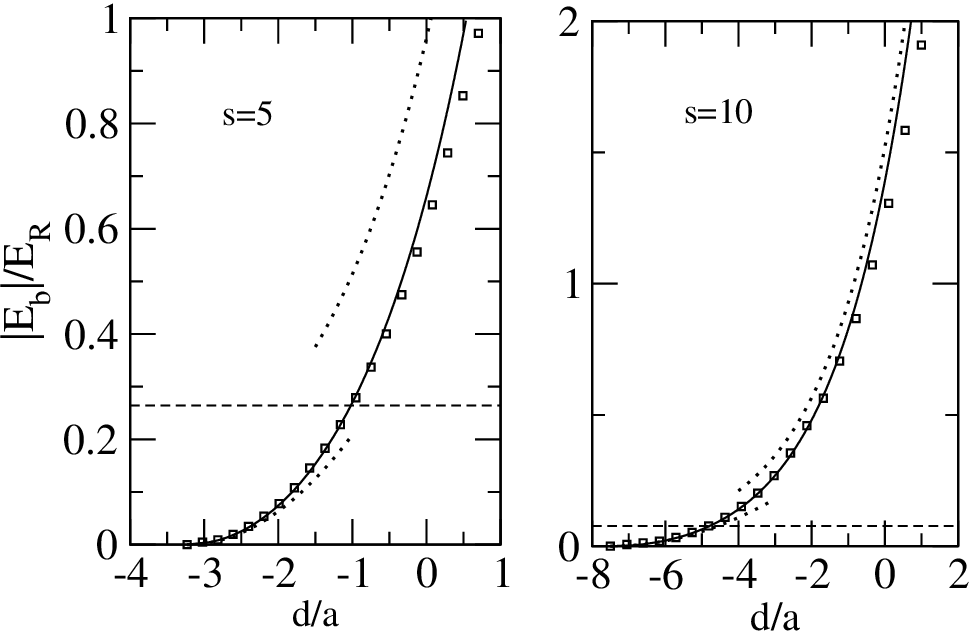}
\caption{Binding energy as a function of the inverse scattering  
length for different values of the laser intensity: the analytical 
prediction (solid line) is compared with the exact numerical data
(squares). Also shown are the position of the bandwidth (dashed line) 
and the asymptotic behaviours for the binding energy
(dotted lines) in the two regimes [see Eqs (\ref{gen}) and (\ref{ho})]. }
\label{figure1}
\end{center}
\end{figure} 
In Fig.\ref{figure1} we plot the binding energy versus 
inverse scattering length derived from Eqs (\ref{eb-1d}) and (\ref{aba}) for different
values of the laser intensity (the value of the 
parameter $d/a_{cr}$ has been taken from Ref.\cite{orso}).  We find that the analytical predictions (solid line) 
are in perfect agreement with the numerical data (squares) provided the binding energy is small compared
to the interband gap. This condition is no longer satisfied approaching the resonance from the negative side, so 
our method is accurate only for scattering length $a<0$. 

Interestingly enough, the fact that $E_b \simeq E_b^{ho} $ for $E_b \gg 4t$ can be used to derive
an analytical expression for the critical scattering length \cite{comtc}. 
By substituting the expansion  $\alpha_{1D}(E_b,0)\simeq -\ln (|E_b|/2t)$
in Eq.(\ref{eb-1d}) with $E_b$ given by Eq.(\ref{ho}), we find  \cite{tc}
\be\label{cric}
\frac{d}{a_{cr}}(Q=0)=-\frac{d}{\sqrt{2\pi}\sigma}\ln
\left(\frac{\lambda \hbar \omega_0}{2t}\right).
\ee
\begin{figure}[tb]
\begin{center}
\includegraphics[width=\columnwidth,angle=0,clip]{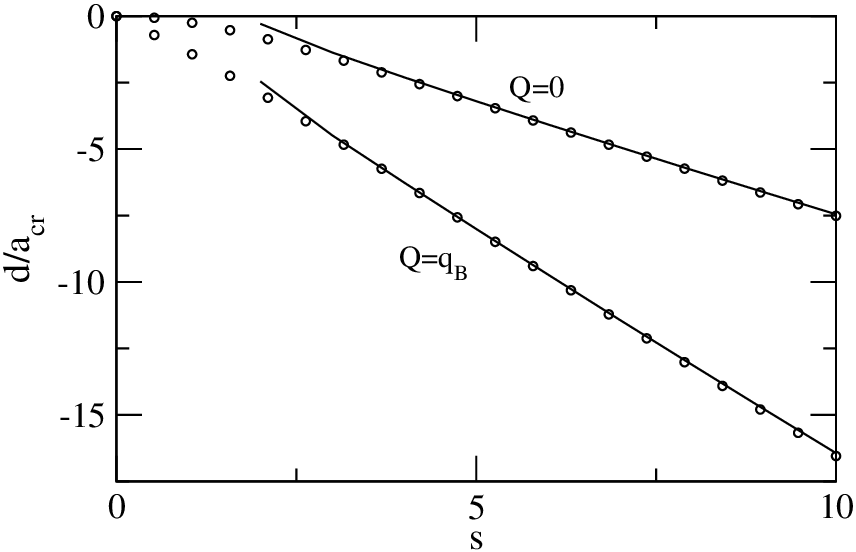}
\caption{Critical value of the inverse scattering length
as a function of the laser intensity for quasi-momentum
$Q=0$ and $Q=q_B$.
The analytical prediction (solid line) is shown with the exact numerical data
(circles).}
\label{figure2}
\end{center}
\end{figure} 
This is plotted in Fig.\ref{figure2} as a function of 
the laser intensity
(solid line) together with the numerical results of Ref.\cite{orso}
(circles). We see that the agreement is excellent already at 
relatively low values of the laser intensity. The approximate linear
behaviour as a function of $s$ shown in Fig.\ref{figure2} is due to the 
logarithmic dependence of the ratio (\ref{cric}) 
on the tunneling rate $t$ and is a specific feature of the case $D=1$.
It is worth pointing out that a good estimate of $a_{cr}$ can be obtained
by the following physical argument. 
In the presence of a deep lattice,
the binding energy is basically given by Eq.(\ref{ho}) provided $|E_b|\gg 4t$.
The tunneling effects that cause the molecule to break become important
when the binding energy (in modulus) is of the order of the bandwidth $4t$
or smaller. This suggest that the critical scattering length $a=a_{cr}$
can be estimated  from Eq.(\ref{ho}) by setting $-E_b^{ho} \simeq 4t$. This 
reproduces the result (\ref{cric}) in the limit $t\ll \hbar \omega_0$.

The effective mass $M^*$ of the molecule is defined as
$\hbar^2/M^*=\partial^2 E(Q)/ \partial Q^2 $ evaluated at $Q=0$.
From Eqs (\ref{eb-1d}-\ref{a-1D}) after some algebra we obtain
\be\label{massa}
\frac{2m^*}{M^*}=1+\frac{|E_b|}{4t}-
\sqrt{\left(\frac{E_b}{4t}\right)^2+\frac{|E_b|}{2t}}.
\ee
where $E_b=E_b(0)$ and $m^*$ is the effective mass for single atoms. We see that
in the anisotropic three dimensional regime, the mass ratio $2m^*/M^*=1$ 
for $E_b=0$ and decreases as $-E_b$ increases.
In the quasi two dimensional regime, $-E_b \gg 4t$, Eq.(\ref{massa}) yields
$2m^*/M^*=2t/|E_b|\ll 1$, showing that the correlation between the two atoms
results in an increased inertia of the molecule. 
\begin{figure}[tb]
\begin{center}
\includegraphics[width=0.85\columnwidth,angle=0,clip]{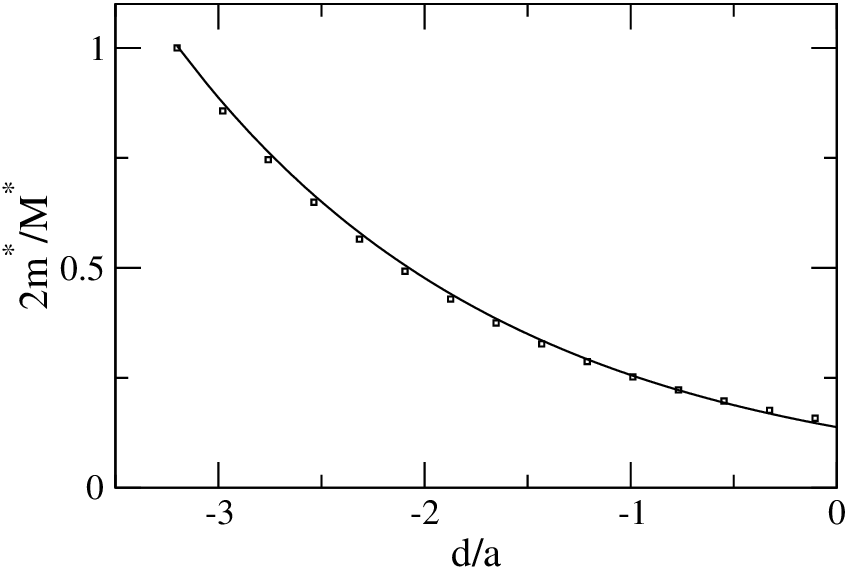}
\caption{The inverse effective mass of the molecule as a function 
of the inverse scattering length for $s=5$: analytical prediction (solid line)
and exact numerical data (squares).} 
\label{figure3}
\end{center}
\end{figure} 
The mass ratio (\ref{massa}) is plotted in Fig.\ref{figure3} against the 
inverse scattering length for $s=5$ togheter with the exact numerical 
result (dots).
We emphasize that the dependence of the effective mass of the molecule
on the value of the scattering length is a clear consequence of the non separability of
the center of mass and the relative motion in the lattice. 

Let us now discuss the solution of Eq.(\ref{int2}) corresponding
to finite value of the quasi-momentum $Q$. 
The binding energy $E_b(Q)$ can be obtained from  Eqs (\ref{eb-1d}) 
and (\ref{a-1D}) provided the critical scattering length $a_{cr}(Q)$ 
is known. This quantity can be easily calculated by noticing that in the
quasi two dimensional limit $|E_b(Q)| \gg 4t$, to lowest order in tunneling,
the binding energy does not depend on the center of mass motion and 
therefore $E_b(Q)=E_b(0)=E_b^{ho}$. 
We have found that for $Q \simeq q_B$ the use of the tight-binding 
dispersion $\epsilon_1(q)=2t(1-\cos(qd))$ in Eq.(\ref{a-1D}) gives 
inaccurate results. This dispersion relation predicts indeed
a continuous degeneracy for the lowest energy state with quasi-momentum 
$Q=q_B$ of the two atoms in the absence of interaction. 
As a result, the integration over quasi-momentum in Eq.(\ref{a-1D}) 
is divergent at this point. This artefact can be easily eliminated by 
using the numerically exact dispersion or by including higher harmonics. 
With this important remark, the  
calculated critical value of the scattering length for 
$Q=q_B$ agrees fairly well with numerics, see Fig.\ref{figure2}. 

\begin{figure}[tb]
\begin{center}
\includegraphics[width=\columnwidth,angle=0,clip]{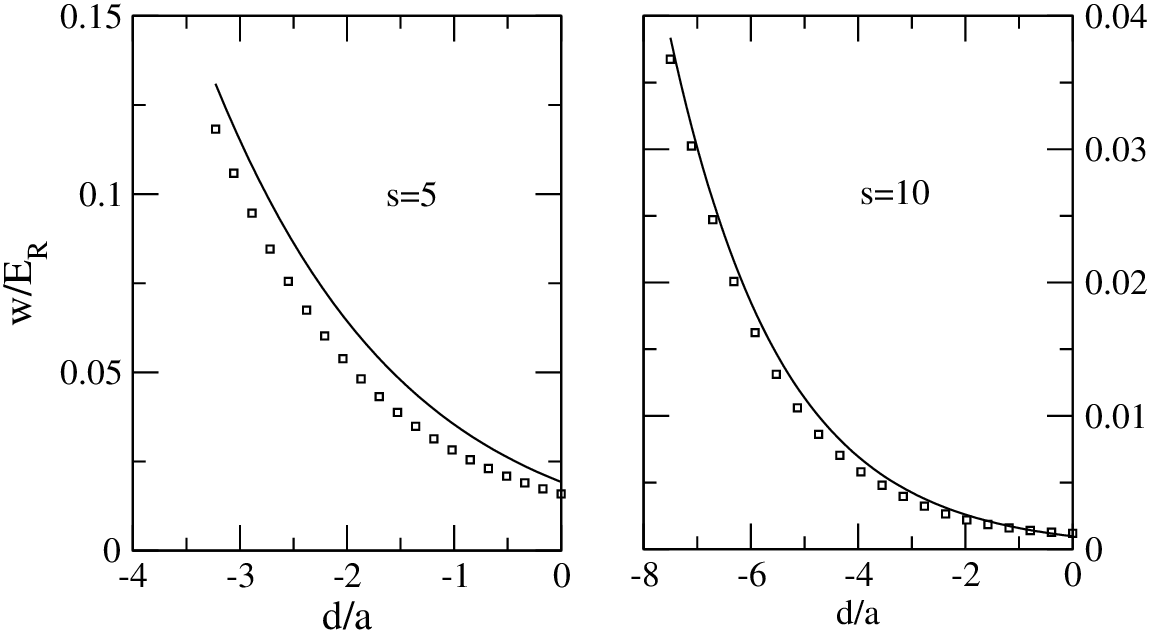}
\caption{Molecular bandwidth as a function of the inverse scattering length
for different values of the laser intensity: analytical prediction 
(solid line) and exact numerical data (squares). The atomic bandwidth is 
$w_{at}=0.264 E_R$ for $s=5$ and $w_{at}=0.077 E_R$ for $s=10$.} 
\label{figure4}
\end{center}
\end{figure}

Another quantity of physical interest is the molecular bandwidth $w=E(Q=q_B)-E(0)$. Within our 
notation, the latter can be written as $w=E_{ref}(q_B)+E_b(q_B)-E_b(0)$.
This is plotted in Fig.\ref{figure4} as a function of the inverse scattering length.
Similarly to the effective mass, the bandwith decreases as $d/a$ increases.
However, for $s=5$ (left panel) there is a discrepancy between the analytical prediction 
and the numerical
result which was not found in Fig.\ref{figure3}. This is due to the fact that 
the interband gap is smaller at the edge of the Brillouin zone and there is a residual contribution
to $E_b(q_B)$ coming from higher bands. This effect disappears when 
$s$ is increased so that for $s=10$ good agreement
with the numerical results is obtained (right pannel). Notice that, 
the molecular bandwidth is always smaller 
than the atomic bandwidth $w_{at}$. This can be seen from the fact that $E_b(q_B)-E_b(0)<0$ 
and $E_{ref}(q_B)=2\epsilon_1(q_B/2) \leq \epsilon_1(q_B)=w_{at}$.  

\subsection{Scattering amplitude}
We assume that the two scattering atoms 
belong to the lowest Bloch band, so the incident state 
in Eq.(\ref{nonomo}) is given by $\Psi_0(\mathbf r,Z)=
\phi_{1 q_1}(z_1)\phi_{1 q_2}(z_2)e^{i \mathbf k_\perp \mathbf r_\perp}$.
The energy is $E=\hbar^2 k_\perp^2/m+\epsilon_1(q_1)+\epsilon_1(q_2)$
and the quasi-momentum $Q$ is given by 
\begin{eqnarray}\label{q}
&Q&=q_1+q_2\;\;\;\;\;\;\; \textrm{if}\;\;\; q_1+q_2 \le q_B\nonumber\\
&& =q_1+q_2\pm 2q_B\;\; \textrm{if}\;\;\; |q_1+q_2|>q_B.
\end{eqnarray}
Notice that we only require $E\ll \epsilon_g$,
but there is no restriction on the value of the scattering length $a$.
We insert the ansatz $f_Q(Z)=A \sum_{\ell} w^2(Z-\ell d)e^{iQZ}$
in Eq.(\ref{nonomo}) where $A$ is a numerical coefficient which
is related to the scattering amplitude (\ref{scatt}) by
$f_{sc}=aCA$, where $C=d/\sqrt{2\pi}\sigma$.
 Repeating the same steps as in the previous section,  
after integration over $Z^\prime$, we find 
\be\label{a}
f_{sc}=\frac{a C}{1-a/a_{cr}+(a/\sqrt{2\pi}\sigma) \beta_{1D}(E,Q)},
\ee
where 
\begin{eqnarray}\nonumber
\beta_{1D}(E,Q)&=&-\int_{-\pi}^{\pi}\frac{dq}{2\pi}  \ln \Big| 
\frac{-E+\epsilon_1(q)+\epsilon_1(Q-q)}{\epsilon_1(q)+\epsilon_1(Q-q)}
\Big|\\
&+&i \pi \int_{-\pi}^{\pi}\frac{dq}{2\pi} 
\Theta(E-\epsilon_1(q)-\epsilon_1(Q-q)).
\label{b1}
\end{eqnarray}
This function is related to the analytical continuation of
$\alpha_{1D}$, defined in Eq.(\ref{a-1D}).
Here $\Theta(x)$ is the unit-step function and we have used 
the formula $(x-E-i0)^{-1}=\textrm{P}(x-E)^{-1}+i\pi \delta(x-E)$,
where the symbol P stands for Principal value.
In the special case $Q=0$, the integration over quasi-momentum 
in Eq.(\ref{b1}) can be done analytically using the  
dispersion relation $\epsilon_1(q_z)=2t(1-\cos(q_z d))$ and 
one finds \cite{tc}  
\begin{eqnarray}\label{bb1} 
&\beta_{1D}(E,0)&=i \arccos(1-E/4t)\;\; \textrm{if $E<8t$},\\
&&-\ln[E(1+\sqrt{1-8t/E})^2/8t]+i\pi\;\;  \textrm{if $E>8t$}.\nonumber
\end{eqnarray}
Equations (\ref{b1}) and (\ref{bb1}) show that the scattering amplitude 
undergoes a dimensional crossover as the ratio $E/8t$ is changed. 
In the anisotropic three dimensional regime $E \ll 8t$, 
one has $f_{sc}=aC/(1-a/a_{cr}+iaC\sqrt{Em^*}/\hbar)$. 
Taking into account Eq.(\ref{cric}), in the quasi two dimensional 
regime $E \gg 8t$  we obtain
$f_{sc}=d(a/\sqrt{2\pi}\sigma)/\left(1+(a/\sqrt{2\pi}\sigma)
(\ln[\lambda \hbar \omega_{0}/E]+i \pi)\right)$ 
and $\lambda=0.915/\pi$ in agreement with Ref.\cite{gora}.

It is interesting to notice that the scattering amplitude 
for $Q=0$ is related to the effective coupling constant
$g_{eff}$ for Fermi atoms undergoing Cooper pairing in the case of weak 
attractive interaction $(a<0)$. 
This result was used in Ref.\cite{tc}
to calculate the superfluid transition temperature 
of a two-component Fermi gas in the presence of a one dimensional 
optical lattice. 

\section{TWO AND THREE DIMENSIONAL LATTICES}
In this section we extend our analysis to higher dimensional lattices.
Differently from the case $D=1$, the integration over $k_\perp$
in Eq.(\ref{GEexplicit}) for $D=2,3$ is convergent for $r_\perp=0$,
which considerably simplifies our analysis. Moreover,
in the limit $E\ll \hbar \omega_0$, the energy dependence of the kernel
comes entirely from the contribution of the lowest band $(\mathbf n_1=\mathbf n_2=\mathbf 1)$. 
The inclusion of higher bands gives small correction to the binding energy but
is crucial to reproduce the correct behaviour of the scattering amplitude close to
Feshbach resonance, as discussed below.

\subsection{Bound state}
For a $D$ dimensional optical lattice, the tight binding ansatz (\ref{ansatz}) for the function 
$f(\mathbf R)$ in Eq.(\ref{int2})  generalizes to
\be\label{genas}
f(\mathbf R)=\Pi_{i=1}^{D}f_{Q_i}(R_i), 
\ee
where $Q_i$ and $R_i$ are the $i$-th components of the quasi-momentum 
$\mathbf Q$ and the center of mass position $\mathbf R$, respectively. 
By inserting the ansatz (\ref{genas}) in Eq.(\ref{int2}) and 
integrating over $\mathbf R^\prime$, we find 
\be\label{b}
1/g=\Lambda_D(E_b,\mathbf Q)+B_D(\mathbf Q),
\ee
where $\Lambda_D$ gives the contribution of the lowest Bloch band
\begin{widetext}
\begin{eqnarray}
  \Lambda_2(E_b,\mathbf{Q})&=&\frac{-C^2\sqrt{m}}{2\hbar} 
\int_{-q_B}^{q_B}
\frac{d^2\mathbf{q}}{(2\pi)^2}\frac{1}{\sqrt{-E_b-2\epsilon_\mathbf 1 (\mathbf Q/2)+\epsilon_{1}(\mathbf{q})+\epsilon_{1}(\mathbf{Q}-\mathbf{q})}}  
  \label{lambda2},   \\
  \Lambda_3(E_b,\mathbf{Q})&=&-C^3 \int_{-q_B}^{q_B} 
  \frac{d^3\mathbf{q}}{(2\pi)^3}\frac{1}{-E_b-2\epsilon_\mathbf 1 (\mathbf Q/2)+\epsilon_{1}(\mathbf{q})+\epsilon_{1}(\mathbf{Q}-\mathbf{q})} \label{lambda3}
\end{eqnarray}
\end{widetext}
and $C$ has been defined previously. The quantity $B_D(Q)$ in Eq.(\ref{b}) takes into account the small correction
coming from the contribution of all higher bands and can be evaluated either numerically or analytically, as explained below. 
Eqs (\ref{b}-\ref{lambda3}) permit us to calculate the binding 
energy as a function of the scattering length for a given quasi-momentum 
$\mathbf Q$. 

For simplicity we restrict our discussion to the case 
$\mathbf Q=0$. In analogy with the case $D=1$, there are two regimes:
an anisotropic three-dimensional regime for $E_b \ll t$ and 
a confined regime of reduced dimensionality 
for $t \ll |E_b|\ll \hbar \omega_0 \sim \epsilon_g$. The constant $B_D$
 can be fixed by matching the binding energy in the limit
$|E_b|\gg t$ with the corresponding result valid for the harmonic oscillator 
potential \cite{busch,olshanii} 
in the limit $|E_b|\ll\omega_0$:
\begin{eqnarray}\label{ols}
|E_b|&=&\frac{\hbar^2}{m}\frac{a^2}{\sigma^4\left(1- 
\eta_2 \frac{a}{\sigma}\right)^2}, \hspace{1cm} \text{(D=2)}\\
|E_b|&=&\frac{\hbar^2}{m}\sqrt{\frac{2}{\pi}}\frac{a}{\sigma^3
\left(1-\eta_3 \frac{a}{\sigma}\right)} \hspace{0.8cm} \text{(D=3)}, \label{busch}
\end{eqnarray}
where $\eta_2=1.0326$ and $\eta_3=0.2448$. This gives $B_D=\eta_D m/4\pi\hbar^2 \sigma$.
Taking into account that the range of applicability of our method is 
$|E_B| \ll \hbar \omega_0$ or, equivalently, $|a|\ll \sigma$, we see that the 
contribution from higher bands to the binding energy is typically small.  

The critical value of the scattering length $a=a_{cr}$ can be 
obtained by setting $E_b=0$ in Eqs (\ref{lambda2}) and (\ref{lambda3})
and integrating over the quasi-momenta using the tight-binding dispersion
$\epsilon_{n}(\mathbf q)=\sum_{j=1}^D 2t(1-\cos(q_jd))$. This gives
\begin{eqnarray}
  \frac{1}{a_{cr}} &=& -\frac{1}{\sigma}\left(I_2 \sqrt{\frac{\hbar \omega_0}{t}}-\eta_2 \right), \;\;\;(D=2) \label{cri-2d}\\
  \frac{1}{a_{cr}} &=& -\frac{1}{\sigma}\left(I_3 \frac{\hbar \omega_0}{t}-\eta_3 \right), \;\;\;\;(D=3)     \label{cri-3d},
\end{eqnarray}
where $I_{2}=0.454$ and $I_{3}=0.101$. Similar results for $D=3$ have 
been obtained by Fedichev et al.\cite{fedichev}  using a different method. 
Similarly to the 1D case, the functional dependence of the critical scattering length
on the ratio $\hbar \omega_0/t$ simply follows from Eqs (\ref{ols}) [D=2] and (\ref{busch}) [D=3]
by substituting the binding energy with the tunneling rate $(|E_b|\rightarrow t)$. 
We see from Eqs (\ref{cri-2d}) and (\ref{cri-3d}) that
 the ratio $d/a_{cr}$ 
has a power law dependence on the tunneling rate $t$ and 
therefore it increases exponentially fast as a function of the 
laser intensity, in contrast with the case $D=1$ shown in Fig.\ref{figure2}.

From Eqs (\ref{lambda2}) and (\ref{lambda3}), we recover the behaviour
of the binding energy close to the critical point
\be\label{uni}
\frac{1}{a}-\frac{1}{a_{cr}}=\frac{1}{\hbar}\frac{m^{*D/2}}{m^{(D-1)/2}}C^D\sqrt{|E_b|},
\ee
which characterizes the three dimensional regime. 

The effective mass $M^*$ for the molecule (at $Q=0$) can be calculated from 
Eqs (\ref{cri-2d}) and (\ref{cri-3d}).  For zero binding energy, 
the mass ratio $2m^*/M^*=1$ and decreases rapidly by increasing $|E_b|$ as an effect 
of the correlated motion of the two atoms.
In the confined regime $|E_b|\gg t$, we find
$2m^*/M^*=(D+1)t/|E_b|$. 

\subsection{Scattering amplitude} 
 We assume again that the two scattering atoms 
belong to the lowest Bloch band, so the incident state in Eq.(\ref{nonomo}) is given by 
$\Psi_0(\mathbf r,\mathbf R)=\phi_\mathbf{1 q_1}(\mathbf x_1)
\phi_\mathbf{1 q_2}(\mathbf x_2)e^{i \mathbf k_\perp \mathbf r_\perp}$
and its energy is $E=\hbar^2 k_\perp^2/m+\epsilon_\mathbf 1 (\mathbf q_1)+
\epsilon_\mathbf 1 (\mathbf q_2)$.
By inserting the ansazt $f_\mathbf Q (\mathbf R)=A\Pi_{i=1}^{D}f_{Q_i}(R_i)$,
in Eq.(\ref{nonomo}) and taking into account that $f_{sc}=aC^D A$, we find
\be\label{a2}
f_{sc}=\frac{a C^D}{1-g (\beta_D(E,\mathbf Q)+B_D)},
\ee 
where $B_D$ has been defined above and
\begin{widetext}
\begin{eqnarray}
\label{beta2}
\beta_2(E,\mathbf Q)&=&\frac{-C^2\sqrt{m}}{2\hbar} 
\int_{-q_B}^{q_B}
\frac{d^2\mathbf{q}}{(2\pi)^2}\left[\frac{\Theta\left(\epsilon_{1}(\mathbf{q})+\epsilon_{1}(\mathbf{Q}-\mathbf{q})-E \right)}
{\sqrt{\epsilon_{1}(\mathbf{q})+\epsilon_{1}(\mathbf{Q}-\mathbf{q})-E}}+i  
\frac{\Theta\left(E-\epsilon_{1}(\mathbf{q})-\epsilon_{1}(\mathbf{Q}-\mathbf{q})\right)}
{\sqrt{E-\epsilon_{1}(\mathbf{q})-\epsilon_{1}(\mathbf{Q}-\mathbf{q})}}
\right],\\
\beta_3(E,\mathbf Q)&=&-C^3 \int_{-q_B}^{q_B} 
  \frac{d^3\mathbf{q}}{(2\pi)^3}\frac{1}{-E-i0+ 
(\mathbf Q/2)+\epsilon_{1}(\mathbf{q})+\epsilon_{1}(\mathbf{Q}-\mathbf{q})}. 
\label{beta3}
\end{eqnarray}
\end{widetext}
Equations (\ref{a2}) and (\ref{beta3}) permit us to calculate the low energy 
scattering amplitude $f_{sc}=aC^2 A$ in the presence of the lattice.
Notice that $f_{sc}$ has a pole at the critical value of the scattering length 
$a=a_{cr}(\mathbf Q)$ for $E=E_{ref}(\mathbf Q)$, corresponding to the bound
state with vanishing binding energy.

For simplicity we confine ourselves to the case $Q=0$ where $E_{ref}=0$.
In the three dimensional regime, from Eqs (\ref{beta2}) and (\ref{beta3}) 
we find the typical result
\be
f_{sc}=\frac{aC^D}{1-a/a_{cr}+i a C^D m^{*D/2} \sqrt{E}/m^{(D-1)/2}\hbar}
\ee
valid in the limit $E\ll t$. This result was also found in Ref. \cite{fedichev} for the case of a 
3D lattice.  In the case $D=2$ the system 
undergoes for $E>8t$ a crossover to a quasi-one dimensional regime. 
The dispersion of the lowest band can be neglected in Eq.(\ref{beta2}) and the
scattering amplitude (\ref{scatt}) reduces to the asymptotic value \cite{olshanii}
\be\label{olsca}
f_{sc} \approx \frac{a C^2 \sqrt{E}}
{(1-\eta_2 \frac{a}{\sigma})\sqrt{E}+i\frac{\hbar a}{\sqrt{m}\sigma^2}}.
\ee
The inclusion of higher bands ($\eta_2 \neq 0$)
becomes crucial when the scattering length is large and only when they are taken 
into account, one recovers the confinement induced resonance for $a=\sigma/\eta_2$.
Notice that our definition of the scattering amplitude [see Eq.(\ref{scatt})] is related
to the amplitude $f(E)$ used in Ref.\cite{olshanii} by 
a scaling factor $f(E)=-i (\sqrt{2\pi}/d^2\sqrt{Em}) f_{sc}(E)$. 

Finally, we have verified that for $D\geq 2$, Eqns (\ref{b}) and (\ref{a2}) 
also follow from a Hubbard-like model with hopping along the
the optical lattice, free motion in the perpendicular direction and
the contact interaction strength equal to 
$U=g/[(1-\eta_Da/\sigma)(\sqrt{2\pi}\sigma)^D]$ \cite{hub1D}.

\section{CONCLUSIONS}
In this paper we have generalized the two-body 
scattering theory to include the effect of periodic fields,  
where the center of mass and the relative motion do not separate. 
The interplay between confinement and tunneling effects gives rise
to an interesting crossover between 
a low dimensional regime where the interwell hopping can be 
treated perturbatively and
an anisotropic three dimensional regime where the wave function
for the relative motion
spreads over many lattice sites.
By using a tight-binding approach, we have investigated the low energy properties 
of the bound state and the scattering amplitude across the two regimes.
The precision of our analytical predictions have been tested
for the case of a one dimensional lattice where exact numerics is available,
finding very good agreement. Our results are relevant for current experiments
on ultracold gases in optical lattices.

We acknowledge interesting discussions with Z. Idziaszek
and thank S. Stringari for comments on the manuscript.
This work was supported by the Ministero dell'Istruzione, 
dell'Universita' e della Ricerca (M.I.U.R.), by 
the Special Research Fund of the University of Antwerp, BOF NOI UA 2004, 
and by the FWO-V project No.G.0435.03. M.W. is financially 
supported by the `FWO -- Vlaanderen'.


\begin{thebibliography}{15}
\bibitem{ketterle} S. Inouye et al., Nature {\bf 392}, 151 (1998). 
\bibitem{varenna} For a review of optical lattice see for instance 
P.S. Jessen and  I.H. Deutsch, Adv. At. Mol. Opt. Phys. {\bf 37}, 95 (1996); 
G. Grynberg and C. Robilliard, Phys. Rep.
{\bf 355}, 335 (2001).
\bibitem{jila} M. Greiner, C. Regal, and D.S. Jin, Nature {\bf 426}, 537 (2003).
\bibitem{Innsbruck} S. Jochim {\sl et al.}, Phys. Rev. Lett. {\bf 91}, 240402 (2003).
\bibitem{mit} M. W. Zwierlein {\sl et al.}, Phys. Rev. Lett. {\bf 91}, 250401 (2003).
\bibitem{ens} J. Cubizolles {\sl et al.}, Phys. Rev. Lett. {\bf 91}, 240401 (2003).
\bibitem{esslinger} H. Moritz {\sl et al.}, Phys. Rev. Lett. {\bf 94}, 210401 (2005).
\bibitem{busch} T. Busch {\sl et al.}, Found. Phys. {\bf 28}, 549 (1998).
\bibitem{olshanii} M. Olshanii, Phys. Rev. Lett. {\bf 81}, 938 (1998), 
T. Bergeman, M. G. Moore and M. Olshanii, Phys. Rev. Lett {\bf 91}, 163201 (2003).
\bibitem{gora} D.S. Petrov, M.Holzmann and G.V. Shlyapnikov, Phys.
Rev. Lett. {\bf 84}, 2551 (2000); D.S. Petrov and G.V. Shlyapnikov, Phys.
Rev. A {\bf 64}, 012706 (2000).
\bibitem{sbiscek} Z. Idziaszek and T. Calarco, Phys. Rev. A {\bf 71}, 
050701(R) (2005).
\bibitem{granger} B.E. Granger, D. Blume, Phys. Rev. Lett. {\bf 92}, 133202 (2004).
\bibitem{sbiscekp}  Z. Idziaszek and T. Calarco, quant-ph/0507186.
\bibitem{orso} G. Orso {\sl et al.}, Phys. Rev. Lett. {\bf 95}, 060402 (2005).
\bibitem{mora} V. Peano {\sl et al.} cond-mat/0506272.
\bibitem{fedichev}P. O. Fedichev, M. J. Bijlsma, and P. Zoller, 
Phys. Rev. Lett. \textbf{92}, 080401 (2004).
\bibitem{tc} G. Orso and G.V. Shlyapnikov, cond-mat/0507597.
\bibitem{note3D} For one and two-dimensional optical lattices, localised solutions exist 
only with $E<E_{ref}$, because the spectrum of two non interacting atoms 
forms a 
continuum for $E>E_{ref}$. For a sufficiently tight
three-dimensional optical lattice, the two particle 
spectrum develops gaps leading to localised solutions with $E>E_{ref}$.
\bibitem{comtc} In Ref. \cite{tc}, the critical value of the scattering length
was derived by comparing the scattering amplitudes. The two methods are 
equivalent.
\bibitem{hub1D} For the one-dimensional optical lattice, the contact 
interaction needs an appropriate regularisation  as discussed in, e.g.
K. W\'{o}dkiewicz, Phys. Rev. A {\bf 43}, 68 (1990).
\end{thebibliography}
\end{document}